\documentclass[12pt]{article}
\usepackage{amsmath,amssymb}
\newcommand{\be}{\begin{equation}}
\newcommand{\ee}{\end{equation}}
\newcommand{\bea}{\begin{eqnarray}}
\newcommand{\eea}{\end{eqnarray}}
\newcommand{\nn}{\nonumber \\}
\newcommand{\p}[1]{(\ref{#1})}
\newcommand{\lb}{\label}

\begin{document}

\begin{center}
{\Large\bf New Approach to Duality-Invariant Nonlinear\\
\vspace{0.2cm}

Electrodynamics} \vspace{0.8cm}

{\large\bf Evgeny Ivanov}$^{\,a)}$, {\large\bf Olaf Lechtenfeld}$^{\,b)}$, {\large\bf Boris Zupnik}$^{\,a)}$
\vspace{0.4cm}

${\,}^{\,a)}\,${\it Bogoliubov Laboratory of Theoretical Physics,
JINR, \\
141980, Dubna, Moscow Region, Russia}\\
{\tt eivanov,zupnik@theor.jinr.ru}\\[8pt]

${\,}^{\,b)}\,${\it Institut f\"ur Theoretische Physik
and Riemann Center for Geometry and Physics, \\
Leibniz Universit\"at Hannover, \\
Appelstra{\ss}e 2, 30167 Hannover, Germany
}\\
{\tt lechtenf@itp.uni-hannover.de}

\end{center}
\vspace{0.5cm}


\begin{center}
{\it Talk at the conference ``Integrable Systems and Quantum Symmetries'', \break Prague, June 11 - 16, 2013}
\end{center}
\vspace{0.5cm}

\begin{abstract}
\noindent
We survey a new approach to the duality-invariant systems of nonlinear electrodynamics, based on introducing
auxiliary bi-spinor fields. In this approach, the entire information about the given self-dual system is encoded in the $U(1)$
invariant interaction of the auxiliary fields, while the standard self-dual Lagrangians appear on shell as a result of eliminating
auxiliary fields by their equations of motion. Starting from the simplest $U(1)$ duality, we show how this approach can be
generalized to the $U(N)$ duality (with $N$ independent Maxwell field strengths), as well as to self-dual systems of ${\cal N}=1$ supersymmetric
electrodynamics. Also, it works perfectly  for self-dual systems with higher derivatives in the action.
\end{abstract}

\section{Motivations}
$U(N)$ duality invariance is on-shell symmetry of a wide class  of the nonlinear electrodynamics models including the renowned
Born-Infeld theory \cite{GZ}-\cite{AFZ}. It is a generalization of the well known $O(2)$ symmetry between the free Maxwell equations
of motion (EOM) and Bianchi identities for
$F_{mn} = \partial_m A_n - \partial_n A_m, \tilde{F}_{mn} = \frac 12 \epsilon_{mnpq}F^{pq}$:
\bea
&&{\rm EOM}: \; \partial^m F_{mn} = 0   \qquad \Longleftrightarrow  \qquad {\rm Bianchi}: \;
\partial^m \tilde{F}_{mn} = 0\,, \lb{free} \\
&& \delta F_{mn} = -\omega \tilde{F}_{mn}, \qquad \delta\tilde{F}_{mn} = \omega F_{mn}\,. \lb{free1}
\eea

In the nonlinear case with the general Lagrangian $L(F)$ the dual field strength is defined  as $P_{mn} = -2\frac{\partial L(F)}{\partial F^{mn}}$,
and eqs. \p{free}, \p{free1} generalize to
\bea
&& {\rm EOM}: \;\partial^m P_{mn} = 0 \qquad \Longleftrightarrow  \qquad
{\rm Bianchi}: \; \partial^m\tilde{F}_{mn} = 0\,, \lb{nonl} \\
&& \delta P_{mn} = -\omega \tilde{F}_{mn}\,, \qquad \delta \tilde{F}_{mn} = \omega {P}_{mn}\,. \lb{nonl1}
\eea
Since $P_{mn}$ is a function of $F_{mn}$, the self-consistency condition (the so called self-duality condition \cite{GZ,GR}) should
be also imposed:
\be
P\tilde{P} + F\tilde{F} = 0\,. \lb{GZ}
\ee
In the case of $N$ Maxwell strengths the $O(2)$ ($U(1)$) duality group is enhanced to $U(N)$.

Recently, there was a rebirth of interest in the duality-invariant theories and their superextensions \cite{BHN}-\cite{AFer},
mainly due to the hypothesis that the generalized duality symmetries at quantum level can play the decisive role
in proving the conjectured UV finiteness of ${\cal N}=8, 4D$ supergravity. Keeping this in mind, new efficient general methods
for treating duality invariant theories and their supersymmetric extensions  are urgently needed.

A decade ago, two of us proposed a new general formulation of the duality invariant theories which
exploits the tensorial (bispinor) auxiliary fields \cite{4}-\cite{6}.
In this approach, the $U(N)$ duality is realized as the
{\it linear} off-shell symmetry of the nonlinear interaction constructed out of the auxiliary fields.
The self-duality constraint \p{GZ} is also {\it linearized}.
In the $U(1)$ case the problem of restoring the nonlinear electrodynamics action by the auxiliary interaction is reduced to solving some
{\it algebraic} equations. On the contrary, in the standard approach,  while restoring the Lagrangian as a solution of the constraint \p{GZ},
one needs to solve {\it differential} equations.

We realized that some methods recently invented for the systematic construction
of various duality invariant systems in \cite{BN}-\cite{CKO} are in fact completely equivalent to our 10-years old approach just mentioned.
This motivated us to return to the original formulation in order to see how the latest developments in the
duality area can be ascribed into its framework \cite{7,9}. Supersymmetrization of the approach with auxiliary bispinor fields was
undertaken in \cite{Ku,8}.\\

The present talk is a brief comprehensive review of all these issues.

\setcounter{equation}{0}
\section{The standard setting}
 We will make use of the bispinor formalism:
\bea
&& F_{mn} \Rightarrow (F_{\alpha\beta}, \quad \bar F_{\dot\alpha\dot\beta}), \quad \varphi := F^{\alpha\beta}F_{\alpha\beta}\,,\quad
\bar\varphi := \bar F^{\dot\alpha\dot\beta}\bar F_{\dot\alpha\dot\beta}\,, \nn
&& L(\varphi,\bar\varphi)=-\frac12(\varphi+\bar\varphi)+L^{int}(\varphi,\bar\varphi) \lb{bisp}
\eea
In this notation, the EOM and Bianchi identities read:
\bea
\partial_\alpha^{\dot\beta} \bar{P}_{\dot\alpha\dot\beta}(F)
-\partial^\beta_{\dot\alpha} P_{\alpha\beta}(F)= 0\,, \quad \partial_\alpha^{\dot\beta} \bar{F}_{\dot\alpha\dot\beta}
-\partial^\beta_{\dot\alpha} F_{\alpha\beta}= 0\,, \quad P_{\alpha\beta}=i\frac{\partial L}{\partial F^{\alpha\beta}}\,.
\eea
The $O(2)$ duality transformations are realized as
\be
\delta_\omega F_{\alpha\beta}=\omega P_{\alpha\beta},\qquad \delta_\omega P_{\alpha\beta}
=-\omega F_{\alpha\beta}\,.
\ee
The self-consistency constraint and the Gaillard-Zumino (GZ) representation for $L$ are:
\bea
&&F_{\alpha\beta}F^{\alpha\beta} + P_{\alpha\beta}P^{\alpha\beta} - \mbox{c.c} = 0 \; \quad
\Leftrightarrow \quad \varphi - \bar\varphi - 4\,[\varphi(L_\varphi)^2 -
\bar\varphi(L_{\bar\varphi})^2]=0\,,\nn
&& L^{sd}=\frac{i}2(\bar{P}\bar{F}-PF)+I(\varphi,\bar\varphi)\,, \quad \delta_\omega I(\varphi,\bar\varphi)=0\,.\lb{GZrepr}
\eea
The GZ representation \p{GZrepr} does not give any explicit recipe how to determine the unknown $U(1)$ duality-invariant function
$I(\varphi, \bar\varphi)$ and so how to restore the whole lagrangian $L^{sd}$. Our approach with the auxiliary tensorial fields provides
an answer to this key question.

\setcounter{equation}{0}
\section{Formulation with bispinor auxiliary fields}
We introduce the auxiliary unconstrained fields $V_{\alpha\beta}$ and $\bar{V}_{\dot\alpha\dot\beta}$
and write the extended Lagrangian  in the $(F,V)$-representation as
\be
{\cal L}(V,F)={\cal L}_2(V,F)+E(\nu,\bar{\nu}), \quad
{\cal L}_2(V,F)=\frac12(\varphi+\bar\varphi)+ \nu+ \bar{\nu} - 2\,(V\cdot F+\bar{V}\cdot\bar{F}), \lb{U1FV}
\ee
where $\nu=V^2,\; \bar\nu=\bar{V}^2\,$. Here ${\cal L}_2(V,F)$ is the bilinear part only through which the Maxwell field strength enters
the action and $E(\nu, \bar\nu)$ is the nonlinear interaction involving only auxiliary fields.

Dynamical equations of motion
\be
\partial_\alpha^{\dot\beta} \bar{P}_{\dot\alpha\dot\beta}(V,F)
-\partial^\beta_{\dot\alpha} P_{\alpha\beta}(V,F)= 0\,, \quad P_{\alpha\beta}(F,V)= i\frac{\partial {\cal L}(V, F)}{\partial F^{\alpha\beta}} =
i(F_{\alpha\beta}-2V_{\alpha\beta}),
\ee
together with Bianchi identity, are covariant under $O(2)$ transformations
\be
\delta V_{\alpha\beta}=-i\omega V_{\alpha\beta}, \;\delta F_{\alpha\beta}=i\omega(F_{\alpha\beta}-2V_{\alpha\beta}),\;\; \delta\nu=-2i\omega\nu\,.
\ee
The algebraic equation of motion for $V_{\alpha\beta}$,
\be
F_{\alpha\beta}=V_{\alpha\beta}+\frac12\frac{\partial E}{\partial V^{\alpha\beta}}
=V_{\alpha\beta}(1+E_\nu)\,,
\ee
is $O(2)$ covariant {\it if and only if} the proper constraint holds:
\be
\nu E_\nu - \bar\nu E_{\bar\nu} = 0 \quad \Rightarrow \quad E(\nu, \bar\nu) = {\cal E}(a)\,, \;\;a := \nu\bar\nu\,.
\ee

The meaning of this constraint is that $E(\nu, \bar\nu)$ should be $O(2)$ invariant function of the auxiliary tensor variables:
\be
\delta_\omega E=2i\omega(\bar\nu E_{\bar\nu}-\nu E_\nu)=0\,.
\ee
This is none other than the self-duality constraint in the new setting:
\be
F^2 + P^2 - \bar{F}^2 - \bar{P}^2 = 0\, \quad \Leftrightarrow \quad \nu E_\nu - \bar\nu E_{\bar\nu} = 0\,.
\ee
The auxiliary equation can be now written as
\be
F_{\alpha\beta}-V_{\alpha\beta} = V_{\alpha\beta}\bar\nu \,{\cal E}'.
\ee
It serves to express $V_{\alpha\beta}, \bar{V}_{\dot\alpha\dot\beta}$ in terms of $F_{\alpha\beta},
\bar{F}_{\dot\alpha\dot\beta}$:
\be
V_{\alpha\beta}(F)=F_{\alpha\beta}G(\varphi,\bar\varphi),\quad G(\varphi,\bar\varphi)=\frac12-L_\varphi=(1+ \bar\nu{\cal E}_a)^{-1}\,.
\ee
After substituting these expressions back into ${\cal L}$ we obtain the corresponding self-dual Lagrangian $L^{sd}(\varphi, \bar\varphi)$
\be
L^{sd}(\varphi, \bar\varphi) = {\cal L}(V(F), F) = -\frac{1}{2}\,\frac{(\varphi + \bar\varphi)
 (1 - a{\cal E}_a^2) + 8a^2{\cal E}_a^3}{1 + a{\cal E}_a^2} + {\cal E}(a)\,,
\ee
where $a$ is related to $\varphi, \bar\varphi$ by the algebraic equation
\be
(1+a{\cal E}_a^2)^2\varphi\bar\varphi=a[(\varphi+\bar\varphi){\cal E}_a
+(1-a{\cal E}_a^2)^2]^2\,.
\ee
The invariant GZ function is now represented as $I(\varphi, \bar\varphi) = {\cal E}(a) - 2a{\cal E}_a$,
$a = \nu(\varphi, \bar\varphi)\bar\nu(\varphi, \bar\varphi)$.

To summarize,  {\it all} $O(2)$ duality-symmetric systems of nonlinear electrodynamics without derivatives
on the field strengths are parametrized by the $O(2)$ invariant off-shell interaction ${\cal E}(a)$ which
is a  function of the real quartic combination of the auxiliary fields. This universality is the basic advantage of the
approach with tensorial auxiliary fields. The problem of constructing $O(2)$
duality-symmetric systems is reduced to choosing one or another specific ${\cal E}(a)\,$.
As distinct from other existing approaches to solving the self-duality constraint, our approach uses
the algebraic equations instead of the differential ones and automatically yields the Lagrangians $L^{sd}(\varphi, \bar\varphi)$
which are analytic at $\varphi = \bar\varphi = 0$.\\

After passing to the tensorial notation, our basic auxiliary field equation (proposed ten years ago) precisely coincides with
what was recently called ``nonlinear twisted self-duality constraint'' \cite{BN}-\cite{CKO},
and ${\cal E}(a)$ with what is called there ``duality invariant source of deformation''.
\setcounter{equation}{0}

\section{Alternative auxiliary field representation}
One of the advantages of our approach is the possibility to choose some other auxiliary
field formulations which sometimes prove to be simpler technically.
It is convenient, e.g.,  to deal with the auxiliary complex variables
$\mu$ and $\bar\mu$ related to $\nu$ and $\bar\nu$ by Legendre transformation
\be
\mu :=E_\nu,\; E-\nu E_\nu-\bar\nu E_{\bar\nu} :=
H(\mu,\bar\mu), \quad  \nu=-H_\mu, \; E = H -\mu H_\mu - \bar\mu H_{\bar\mu}\,.
\ee
Under the $O(2)$ duality $\delta_\omega \mu = 2i\omega\,\mu, \;\delta_\omega \bar\mu = -2i\omega\,\bar\mu \,$, so
$$
H(\mu, \bar\mu) = I(b)\,, \qquad  b = \mu\bar\mu\,.
$$
The basic relations of this formalism are
\bea
&&L^{sd}(\varphi, \bar\varphi) = -\frac{1}{2}\,(\varphi + \bar\varphi + 4bI_b)\,\frac{1-b}{1+b}
+ I(b)\,, \nn
&& (b+1)^2\,\varphi\bar\varphi=b\,[\varphi+\bar\varphi-I_b(b-1)^2]^2\,.
\eea
These relations can be reproduced by eliminating $\mu$ and $\bar\mu$ as independent
scalar auxiliary fields from the off-shell Lagrangian
\be
\tilde{L}(\varphi,\mu)=\frac{\varphi(\mu-1)}{2(1+\mu)}
+\frac{\bar\varphi(\bar\mu-1)}{2(1+\bar\mu)}+I(\mu\bar\mu).
\ee
Yet there even exists a combined tensor - scalar auxiliary field ``master'' formulation which enables to establish relations
between different equivalent off-shell
descriptions of the same duality-invariant system:
\bea
{\cal L}(F,V,\mu)=
\frac{1}2(F^2+\bar{F}^2)-2(VF)-(\bar{V}\bar{F})+V^2(1+\mu)+(1+\bar\mu)\bar{V}^2
+I(\mu\bar\mu)\,.
\eea
\setcounter{equation}{0}

\section{Examples}
\noindent{\bf I. Born - Infeld.}
This model  has a simpler description in the $\mu$ (or $b$) representation:
\be
I^{BI}(b)=\frac{2b}{b-1},\qquad I^{BI}_b=-\frac{2}{(b-1)^2}\,.
\ee
The equation for computing $b$ becomes quadratic:
\bea
&& \varphi\bar\varphi\, b^2+[2\varphi\bar\varphi-(\varphi+\bar\varphi+2)^2]\,b+
\varphi\bar\varphi=0\quad \Rightarrow \nn
&& b = \frac{4\varphi\bar\varphi}{[2(1 + Q) +\varphi + \bar\varphi]^2}\,, \quad
Q(\varphi) := \sqrt{1 + \varphi + \bar\varphi + (1/4)(\varphi - \bar\varphi)^2}\,.
\eea
After substituting this solution into the general formula for $L^{sd}(\varphi, \bar\varphi)$
the standard BI Lagrangian is recovered
\be
L^{BI} (\varphi, \bar\varphi) = 1 - \sqrt{1 + \varphi + \bar\varphi + (1/4)(\varphi - \bar\varphi)^2}\,.
\ee
\vspace{0.3cm}

\noindent{\bf II. The simplest interaction (SI) model.} This model is the simplest example of the
auxiliary interaction generating the non-polynomial self-dual electromagnetic Lagrangian.
The relevant interaction in both $\nu$ and $\mu$ representations are linear functions
\bea
{\cal E}^{SI}(a) = \frac12\, a\,, \quad I^{SI}(b) = -2 b\,, \quad a = \nu\bar\nu, \;\;b = \mu \bar\mu\,.
\eea

Despite such a simple off-shell form of the auxiliary interaction, it is difficult to
find a closed on-shell form of the nonlinear Lagrangian $L^{SI}(\varphi, \bar\varphi)$,
since the algebraic equations relating $a$ (or $b$) to $\varphi, \bar\varphi$ are of the
5-th order. E.g.,
$$
(b +1)^2\,\varphi\bar\varphi = b\,[\varphi + \bar\varphi +2(b-1)^2 ]^2\,.
$$
Nevertheless, it is straightforward to solve these equations as infinite series in  $b$ or $a$ and
then to restore $L^{SI}(\varphi, \bar\varphi)$ to any order.
Up to 10-th order in $F, \bar F$:
\bea
L^{SI}(\varphi, \bar\varphi) &=& -\frac12(\varphi+\bar\varphi)
+e_1\varphi\bar\varphi
-e_1^2(\varphi^2\bar\varphi+\varphi\bar\varphi^2)
+e_1^3(\varphi^3\bar\varphi+\varphi\bar\varphi^3) \nn
&&+\, 4e_1^3 \varphi^2\bar\varphi^2
-e_1^4(\varphi^4\bar\varphi+\varphi\bar\varphi^4)
-10e_1^4(\varphi^3\bar\varphi^2+\varphi^2\bar\varphi^3)+O(F^{12}).
\eea
Here $e_1 = \frac12\,$.
\setcounter{equation}{0}

\section{Systems with higher derivatives}
The nonlinear electromagnetic Lagrangians with higher derivatives are functions
of the variables
$$
F,\quad \partial_m F,\quad \partial_m\partial_n F,
\quad \partial_m\partial_n\partial_r F\ldots
$$
and their complex conjugates.
The higher-derivative Lagrangians
in the explicit form involve various scalar combinations of these variables, e.g.,
$$
F^2, \;\;(\partial^mF\partial_mF), \;\; (\partial^mF^2\partial_mF^2),\quad
(F\Box^N F),\ldots\,.
$$
It is known that the higher-derivative generalizations of the duality-invariant Lagrangians
contain all orders of derivatives of $F_{\alpha\beta}$ and $F_{\dot\alpha\dot\beta}$ \cite{BN}.

In our formulation the generalized self-dual Lagrangian is
\bea
{\cal L}(F,V,\partial V)={\cal L}_2+ {\cal E}(V,\partial V)\,,
\eea
where ${\cal L}_2$ is the same ``free'' bilinear part as before and ${\cal E}(V,\partial V)$ is the $O(2)$
invariant self-interaction which can now involve any Lorentz invariant combinations of the tensorial auxiliary fields and  their derivatives.
In order to avoid non-localities and ghosts, it is reasonable to assume that ${\cal E}(V,\partial V)$ contains no terms bilinear in
$V, \bar V$, i.e. that the extra derivatives appear only at the interaction level.

The equations of motion for this Lagrangian contain the Lagrange
derivative of ${\cal E}$
\bea
\partial^\alpha_{\dot\beta}(F-2V)_{\alpha\beta}+\partial^{\dot\alpha}_{\beta}
(\bar{F}-2\bar{V})_{\dot\alpha\dot\beta} = 0\,, \quad
F_{\alpha\beta}=V_{\alpha\beta}+\frac12\frac{\Delta {\cal E}}
{\Delta V^{\alpha\beta}}\,.
\eea
This set of  equations together with the Bianchi identity for $F, \bar F$ is covariant under the $O(2)$ duality
transformations, provided that ${\cal E}(V, \partial V)$ is $O(2)$ invariant,
$$
\delta_\omega\, {\cal E}(V, \partial V) = 0\,.
$$
An analog of the condition \p{GZ} is the vanishing of the integral
\bea
\int d^4x[P^2(F,V)+F^2-\bar{P}^2(F,V)-\bar{F}^2]=0\,, \; P_{\alpha\beta} = i\frac{\Delta {\cal E}}{\Delta F^{\alpha\beta}} =i(F-2V)_{\alpha\beta}\,,
\eea
and this again amounts to the $O(2)$ invariance of ${\cal E}(V, \partial V)$.

Due to the property that the derivatives appear only in the interaction, one can solve the auxiliary field equations
for $V_{\alpha\beta}, \bar V_{\dot\alpha\dot\beta}$ by recursions, like in the case without derivatives, and to finally
obtain $L^{sd}(F, \partial F)$ as a series expansion to any order in derivatives and the field strengths.\\

\noindent{\bf Some examples.} As the first example we consider
\bea
{\cal E}_{(2)}=\frac12\nu\bar\nu+c\partial^m\nu\partial_m\bar\nu, \quad
\frac{\Delta {\cal E}_{(2)}}{\Delta V^{\alpha\beta}}=2V_{\alpha\beta}[1+\frac12\bar\nu
-c\Box\bar\nu]\,.
\eea
The auxiliary field equation is:
\bea
F_{\alpha\beta}=V_{\alpha\beta}(1+\frac12\bar\nu-c\Box \bar\nu)\,.
\eea
Its perturbative solution is given by:
\bea
V^{(1)}_{\alpha\beta} &=& F_{\alpha\beta}\,,\nn \quad V^{(3)}_{\alpha\beta} &=& -F_{\alpha\beta}(\frac12\bar\varphi -c\Box\bar\varphi),\nn
V^{(5)}_{\alpha\beta} &=& F_{\alpha\beta}\{\frac12\bar\varphi(\frac12\bar\varphi -c\Box\bar\varphi)
+\varphi(\frac12\bar\varphi-c\Box\bar\varphi)-c(\Box \bar\varphi)
(\frac12\bar\varphi-c\Box\bar\varphi)\nn
&&-\,\, 2c\Box[\bar\varphi(\frac12\varphi-c\Box \varphi)]\}, \qquad \mbox{etc}\,,\nonumber
\eea
\bea
L(F,\partial^{N}F) &=& -\frac12(\varphi+\bar\varphi)
+\frac12\varphi\bar\varphi-\frac14\varphi^2\bar\varphi-\frac14\varphi\bar\varphi^2 + c\partial^m\varphi
\partial_m\bar\varphi
\nn
&& +\,c\varphi\bar\varphi[(\Box\varphi)
+(\Box\bar\varphi)]-c^2\varphi(\Box\bar\varphi)^2-c^2\bar\varphi(\Box\varphi)^2+O(F^{8})\,.
\eea
This model can be regarded as the ``minimal'' higher-derivative deformation of the SI model (the latter is recovered at $c=0$).

As the second example, we consider
\bea
{\cal E}_{(4)} = \gamma(\partial^mV\partial^nV)
(\partial_m\bar{V}\partial_n\bar{V}),
\eea
where $\gamma$ is a coupling constant and brackets denote traces with respect to  the $SL(2,C)$ indices.

The auxiliary field equation reads:
\bea
&&F_{\alpha\beta}=V_{\alpha\beta}- \gamma \partial^m\left[
\partial^nV_{\alpha\beta}
(\partial_m\bar{V}\cdot\partial_n\bar{V})\right].
\eea
The perturbative solution is given by:
\bea
&&V^{(1)}_{\alpha\beta}=F_{\alpha\beta},\quad
V^{(3)}_{\alpha\beta}=\gamma\partial^m\left[
\partial^nF_{\alpha\beta}
(\partial_m\bar{F}\cdot\partial_n\bar{F})\right], \quad {\rm etc}\,. \nonumber
\eea
The Lagrangian in the $F$-representation involves
higher derivatives, starting from the sixth order in fields
\bea
L^{(6)}=(V^{(3)}V^{(3)})-2\gamma[(V^{(3)}\partial^n\partial^mF)(\partial_m\bar{F}\partial_n\bar{F})
+ (V^{(3)}\partial^mF)\partial^n(\partial_m\bar{F}\partial_n\bar{F})]
+\mbox{c.c.}\,.
\eea
\setcounter{equation}{0}

\section{$U(N)$ case}
\noindent{\bf The standard setting.} The starting point is the nonlinear Lagrangian with $N$ abelian gauge field strengths
\bea
&& L(F^k,\bar{F}^l)=-\frac12[(F^kF^k)+(\bar{F}^k\bar{F}^k)]+L^{int}(\varphi^{kl}, \bar\varphi^{kl})\,,\lb{Ncase} \\
&&\varphi^{kl}=\varphi^{lk}=(F^kF^l)\,,\quad
\bar\varphi^{kl}=(\bar{F}^k\bar{F}^l)\,.\nonumber
\eea
It is chosen to be invariant off shell under $O(N)$ transformations
\bea
\delta_{\xi} F^k_{\alpha\beta}=\xi^{kl}F^l_{\alpha\beta}\,,\quad
\delta_{\xi}\bar{F}^k_{\dot\alpha\dot\beta}=\xi^{kl}\bar{F}^k_{\dot\alpha\dot\beta}\,,\quad
\xi^{kl}=-\xi^{lk}\,,
\eea
where $\xi^{lk} = -\xi^{kl}$ are the corresponding group parameters.
The nonlinear equations of motion
\bea
&&E^k_{\alpha\dot\alpha} :=\partial_\alpha^{\dot\beta}
\bar{P}^k_{\dot\alpha\dot\beta}(F) -\partial^\beta_{\dot\alpha}
P^k_{\alpha\beta}(F)= 0\,,\; P^k_{\alpha\beta}(F)=i\frac{\partial L}{\partial
F^{k\alpha\beta}}\,, \nonumber
\eea
together with the Bianchi identities
$$
B^k_{\alpha\dot\alpha}=\partial_\alpha^{\dot\beta}
\bar{F}^k_{\dot\alpha\dot\beta} -\partial^\beta_{\dot\alpha}
F^k_{\alpha\beta}= 0\,,
$$
are on-shell covariant under the $U(N)$ duality transformations
\be
\delta_{\eta}
F^k_{\alpha\beta}=\eta^{kl}P^l_{\alpha\beta}\,, \quad \delta_{\eta} P^k_{\alpha\beta}=-\eta^{kl}F^l_{\alpha\beta}\,,\;\eta^{kl}=\eta^{lk}\,,
\ee
where $\eta^{kl}=\eta^{lk}\,$  are $\frac12 N(N+1)$ real parameters completing  $O(N)$ to $U(N)\,$,
provided that the appropriate generalized self-duality consistency conditions hold:
\bea
(P^kP^l)+(F^kF^l)- {\rm c.c.} =0\,, \quad (F^kP^l)-(F^lP^k)-{\rm c.c.} = 0\,.
\eea
\vspace{0.2cm}

\noindent{\bf The $(F, V)$ representation.} An $U(N)$ analog of the Lagrangian \p{U1FV} is \cite{9}
\bea
{\cal L}(F^k,V^k)&=&{\cal L}_2(F^k,V^k)+E(\nu^{kl}, \bar{\nu}^{kl})\,,\;\;\nu^{kl}=(V^kV^l)\,, \quad \bar\nu^{kl}=(\bar V^k\bar V^l)\,,
\\
{\cal L}_2(F^k,V^k) &=& \frac12[(F^kF^k)+(\bar{F}^k\bar{F}^k)]-2[(F^kV^k)
+(\bar{F}^k\bar{V}^k)] \nn
&& +\, (V^kV^k)+(\bar{V}^k\bar{V}^k)\,.
\eea
The $U(N)/O(N)$ duality transformations are implemented as
\bea
\delta_{\eta}
F^k_{\alpha\beta}=\eta^{kl}P^l_{\alpha\beta}=i\eta^{kl}(F^l-2V^l)_{\alpha\beta}\,,\quad
\delta_{\eta} P^k_{\alpha\beta}=-\eta^{kl}F^l_{\alpha\beta}\,.
\eea
For the whole set of equations of motion to be $U(N)$ duality invariant,
the interaction $E(\nu^{kl}, \bar\nu^{kl})$ should be  $U(N)$ invariant:
\bea
E(\nu^{kl},\bar\nu^{kl})~ \Rightarrow~ {\cal E}(A_1, \dots\,, A_N)\,, \qquad  A_1 = \bar\nu^{kl}\nu^{lk}\,, \ldots
\eea
The algebraic equations of the $U(N)$ duality-invariant models  are
\bea
(F^k-V^k)_{\alpha\beta}={\cal E}^{kl}V^l_{\alpha\beta}\,,\quad
(\bar{F}^k-\bar{V}^k)_{\dot\alpha\dot\beta}= \bar{\cal
E}^{kl}\bar{V}^l_{\dot\alpha\dot\beta}\,, \quad {\cal E}^{kl}:= \frac{\partial{\cal E}}{\partial \nu^{kl}}\,.
\eea
These equations of motion are equivalent to the general ``nonlinear twisted
self-duality constraints''. Solving them, e.g.,  by recursions, we can restore the whole nonlinear $U(N)$ duality invariant action
by the invariant interaction ${\cal E}(A_1, \dots\,, A_N)$. The duality invariant actions with higher derivatives can be constructed
by the $U(N)$ invariant interaction involving derivatives of the auxiliary tensorial fields, like in the $U(1)$ case.
\setcounter{equation}{0}

\section{Auxiliary superfields in ${\cal N}{=}1$ electrodynamics}
 The superfield action of nonlinear ${\cal N}=1$ electrodynamics can be written as:
\bea
&& S(W) = \frac1{4}\int d^6\zeta W^2 +\frac1{4}\int d^6\bar\zeta \bar
W^2\, + \frac14 \int d^8 z\,  W^2\bar{W}^2\Lambda(w,\bar{w}, y, \bar{y})\,, \\
&&w = \frac18\bar D^2\bar{W}^2~,\quad \bar{w}=\frac18D^2W^2~,\qquad \quad
y\equiv D^\alpha
W_\alpha = \bar D_{\dot\alpha}\bar{W}^{\dot\alpha}\,. \nonumber
\eea
Here, $W_\alpha(x, \theta_\alpha, \bar\theta_{\dot\beta}) = \frac{i}{2}(\sigma^m\bar\sigma^n)_\alpha^\beta F_{mn}\theta_\beta
- \theta_\alpha D + \ldots\,,$ is the spinor chiral ${\cal N}=1$ Maxwell superfield strength ($\bar D_{\dot\alpha}W_\alpha =0$,
$D^\alpha W_\alpha = \bar D_{\dot\alpha}\bar{W}^{\dot\alpha}$).

The on-shell $U(1)$ duality rotations and the ${\cal N}=1$ analog of the self-duality constraint read \cite{KT}
\bea
&& \delta W_\alpha=\omega
M_\alpha(W,\bar W)~,\quad \delta  M_\alpha=-\omega W_\alpha\, \quad M_\alpha :=-2i\frac{\delta S}{\delta W^\alpha}\,, \\
&& \mbox{Im} \int d^6\zeta\,(W^2 +M^2) = 0\,.
\eea
The ${\cal N}=1$, $U(1)$ duality is the symmetry between the superfield equations of motion and Bianchi
identity \cite{KT}:
\bea
D^\alpha M_\alpha-\bar{D}_{\dot\alpha}
\bar{M}^{\dot\alpha}=0\quad \Leftrightarrow \quad D^\alpha W_\alpha - \bar D_{\dot\alpha}\bar{W}^{\dot\alpha} = 0\,.
\eea
\vspace{0.3cm}

\noindent{\bf How to supersymmetrize the bispinor formulation?} The basic idea \cite{Ku,8} is to embed tensorial auxiliary fields into chiral auxiliary superfields
\bea
V_{\alpha\beta}(x) \,\Rightarrow \, U_\alpha(x, \theta, \bar\theta) = v_\alpha(x) + \theta^\beta V_{\alpha\beta}(x) + \ldots, \quad
\bar D_{\dot\gamma}U_\alpha(x, \theta, \bar\theta) =0
\eea
(similarly, for the ${\cal N}=2$ case).

We make the substitution $S(W) \rightarrow S(W, U)$, with \cite{8}
\bea
S(W, U) &=& \int d^6\zeta\left(U W-{1\over2}U^2-{1\over4}W^2\right)+\mbox{c.c.} \nn
&& +\, \frac{1}{4}\int d^8z\, U^2\bar U^2\,E(u,\bar{u}, g,\bar{g})\,, \lb{N1WU} \\
&& u=\frac18\bar{D}^2\bar{U}^2, \quad \bar{u}=\frac18D^2U^2, \quad
g=D^\alpha U_\alpha\,.\nonumber
\eea
The duality-invariant ${\cal N}=1$ systems amount to the special choice of the interaction in \p{N1WU} as $U(1)$ invariant one
\bea
E_{inv} ={\cal F}(B, A, C)+\bar{\cal F}(\bar{B}, A, C)\,, \quad A := u\bar u\,, \; C :=g\bar{g}\,,\; B := ug^2\,, \;
\bar{B} :=\bar{u}\bar{g^2}\,.
\eea

One can also define the $M$ -representation of the self-dual ${\cal N}=1$ systems as a generalization of the $\mu$ representation
of the bosonic systems. The corresponding superfield action is as follows
\bea
&& S(W,U,M) =  \int d^6\zeta\left(U W-{1\over2}U^2-{1\over4}W^2\right)+\mbox{c.c.} + S_{int}(W, U, M)\,, \\
&& S_{int}(W, U, M) = \frac14 \int d^8z\left[(U^2\bar M + \bar U^2 M) +
M\bar M\,J\,(m, \bar m) \right].
\eea
Here $m = \frac18 \bar D^2 \bar M\,, \; \bar m  = \frac18 D^2 M$ and $M$ is a complex general scalar ${\cal N}=1$ superfield.
The duality transformations are realized as
\bea
\delta M = 2i\omega M\,, \quad \delta \bar M = -2i \omega \bar M\,,\quad
\delta\bar{m}=2i\omega\bar{m}\,,\quad \delta m=2i\omega m\,.
\eea
The duality invariant systems correspond to the choice
$$
J(m, \bar m) = J_{inv}(B)\,, \quad B := m\bar m\,.
$$
\vspace{0.2cm}

\noindent{\bf Example} \cite{8}. ${\cal N}=1$ Born-Infeld theory:
\bea
J_{inv}^{(BI)}(B) = \frac{2}{B - 1}\,.
\eea
This should be compared with the standard $W, \bar W$ representation of the same theory \cite{BIN1}
\bea
&&S_{int}^{(BI)} = \frac14 \int d^8 z\,  W^2\bar{W}^2\Lambda^{(BI)}(w,\bar{w})\,, \quad  w = \frac18\bar D^2\bar{W}^2\,,\; \bar{w}=\frac18D^2W^2\,, \nn
&&\Lambda^{(BI)}(w,\bar{w})=\left[1+{1\over2}(w+\bar{w})+\sqrt{1+(w+\bar{w})
+{1\over4}(w-\bar{w})^2}\right]^{-1}.  \nonumber
\eea
\setcounter{equation}{0}

\section{Summary and outlook}
\begin{itemize}
  \item {\it All} duality invariant systems of nonlinear electrodynamics (including those with higher derivatives) admit an off-shell formulation
  with the auxiliary bispinor (tensorial) fields. These fields are fully unconstrained off shell, there is no need to express them through
  any secondary gauge potentials, etc.

 \item The full information about the given duality invariant system is encoded in the $O(2)$ invariant interaction function
 which depends only on the auxiliary fields (or also on their derivatives) and can be chosen {\it at will}. In many cases it looks much
 simpler compared to the final action written in terms of the Maxwell field strengths.

  \item
  The renowned nonlinear self-duality constraint is {\it linearized} in the new formulation and becomes just the requirement of $O(2)$
  invariance of the auxiliary interaction. The $O(2)$ (and, in fact, $U(N)$ \cite{9})
   duality transformations are linearly realized off shell.

  \item
  The basic algebraic equations eliminating the auxiliary tensor fields are equivalent to the recently employed
  {\it ``nonlinear twisted self-duality constraints''}. In our approach this sort of conditions appear as {\it equations of motion}
  corresponding to the well defined off-shell Lagrangian.
\end{itemize}

\noindent{\bf Some further lines of development}
\begin{itemize}

\item
(a) The full construction and exploration of extensions to ${\cal N}=1$, ${\cal N}=2$, $\ldots$ supersymmetric duality systems,
including the most interesting supersymmetric Born-Infeld theories, in both flat and supergravity backgrounds, as a further development of the
study in \cite{Ku,8}.

\item (b) Adding, in a self-consistent way, scalar and other fields into the auxiliary tensorial field formulation: $U(1)$
duality group $\Rightarrow  SL(2, R)$ \cite{GR}. One adds two scalar fields,
axion and dilaton ${\cal S}_1$ and ${\cal S}_2$, which support a nonlinear realization
of $SL(2,R)$, and properly generalizes the self-dual Lagrangian in the $(F, V)$-representation.
The corresponding equations of motion together with Bianchi identity exhibit $SL(2,R)$ duality invariance. Similarly,
$U(N)$ duality can be extended to $Sp(2N,R)$ via coupling to the coset $Sp(2N,R)/U(N)$ fields. Supersymmetric versions
of this new setting for generalized self-duality can also be constructed.

\item (c) The formulation presented suggests  a  {\it new look} at the duality invariant systems:
in both the {\it classical} and the {\it quantum} cases {\it not} to eliminate
the tensorial auxiliary fields by their equations of motion, but to deal with the off-shell actions {\it at all intermediate steps}.
In many cases, the interaction looks much simpler when  the auxiliary fields are retained in the action. In this connection, recall the off-shell superfield
approach in supersymmetric theories, which in many cases radically
facilitates the quantum calculations and unveils the intrinsic geometric properties of the corresponding theories without
any need to pass on shell by eliminating the auxiliary fields (at least until the final step of calculating $S$ matrix, etc).
It is worth recalling that the tensorial  auxiliary fields have originally appeared  just within an off-shell superfield formulation,
that of ${\cal N}=3$ supersymmetric Born-Infeld theory \cite{4}.
\end{itemize}

\section*{Acknowledgements}

\noindent We acknowledge a support from a grant of the Heisenberg-Landau Program, from the RFBR
grants Nr. 12-02-00517, Nr. 13-02-91330 and the grant DFG LE 838/12-1.
E.I. thanks the organizers of the conference ISQS'2013 for inviting him to present this talk and for kind hospitality in Prague.

\end{document}